\documentclass[nofootinbib,aps,letterpaper,superscriptaddress,showpacs]{revtex4}
\usepackage{amsmath}
\usepackage{amsfonts}
\usepackage[dvips]{graphicx}
\begin{document}

\newcommand\be{\begin{equation}}
\newcommand\ee{\end{equation}}
\newcommand\bea{\begin{eqnarray}}
\newcommand\eea{\end{eqnarray}}
\newcommand\bseq{\begin{subequations}} 
\newcommand\eseq{\end{subequations}}
\newcommand\bcas{\begin{cases}}
\newcommand\ecas{\end{cases}}
\newcommand{\p}{\partial}
\newcommand{\f}{\frac}

\title{Cosmological bounce from a deformed Heisenberg algebra}

\author{Marco Valerio Battisti}
\email{battisti@icra.it}
\affiliation{Dipartimento di Fisica (G9) and ICRA - International Center for Relativistic Astrophysics, Universit\`a di Roma ``Sapienza'' P.le A. Moro 5, 00185 Rome, Italy}


\begin{abstract}
The implications of a deformed Heisenberg algebra on the Friedmann-Robertson-Walker cosmological models are investigated. We consider the Snyder non-commutative space in which the translation group is undeformed and the rotational invariance preserved. When this framework is implemented to one-dimensional systems (which is this case) the modifications are uniquely fixed up to a sign. A cosmological quantum bounce \`a la loop quantum cosmology is then obtained. We also get the Randall-Sundrum braneworld scenario and this way a Snyder-deformed quantum cosmology can be considered as a common phenomenological description for both theories. 
\end{abstract}


\maketitle 

\section{Introduction}

The emergence of space-time singularities in Einstein general relativity undoubtedly implies that such a classical description is valid only on macroscopic scale \cite{HE}. One of the most important example is the big-bang singularity appearing in the standard model of cosmology. However, the Friedmann dynamics is expected to be modified by quantum effects in the regime of small scale factor and that such corrections naturally come out from a quantum theory of gravity. Anyway, it is not yet clear which kind of modifications may appear and this problem is somehow related to the one of finding the right (phenomenological) description of the low energy limit of quantum gravity. In particular, it is argued that this limit should contain the notion of an other invariant, observer independent, scale (the Planck scale) \cite{RovSmo} and this can be regarded as the main intuition of doubly special relativity (DSR) \cite{AMS}. 

In this work the (deformed) dynamics of the Friedmann-Robertson-Walker (FRW) Universe is analyzed in the context of the Snyder non-commutative space \cite{Sny}. The Snyder approach is relevant since it can be related to some of DSR models \cite{Kov} and has some motivations from loop quantum gravity \cite{LO} as well as two-time physics \cite{tt}. 

The only deformed commutator in the Snyder framework is the one between the coordinates, i.e. the translation group is not modified and the rotational symmetry is preserved. This way, infinitely many generalized uncertainty relations, underlying deformed Heisenberg algebras, are predicted \cite{BatMel08}. Assuming the FRW phase space as Snyder-deformed (in this case we deal with a one-dimensional system), the scheme is almost uniquely fixed. A non-singular bouncing cosmological evolution is then predicted. Notably, the modified Friedmann equation resembles the one arising in loop quantum cosmology (LQC) \cite{elqc}. It is worth noting that, since the Snyder picture is fixed up to a sign, also the Randall-Sundrum braneworld scenario \cite{Roy} is predicted by our model. In other words, this deformed phase space can be considered, from a phenomenological point of view, as an effective framework which is able to describe the results obtained in more general theories. From this perspective, the different predictions of such approaches can then be understood considering the opposite sign in the deformation term. 

Our model can be regarded as an important step toward the comparison between deformed and loop quantum cosmology. In fact the quantum behavior of the FRW Universe, using deformed Heisenberg algebras reproducing the string theory uncertainty principle \cite{String}, has been analyzed to describe the fate of the big-bang singularity \cite{BM07}. However, in the previous work the classical singularity appears to be probabilistically suppressed, but no evidences for a big-bounce arise. (For a comparison between deformed and polymer-loop quantum cosmology in the Taub Universe see \cite{BM08}.) On the other hand, we now deal with a more general framework (the previous approach is recovered as a particular case) and a cosmological bounce \`a la loop is then allowed.  

The paper is organized as follows. In Section II the Snyder space and its deformed quantum mechanics are analyzed. Section III is devoted to the study of the modified FRW cosmological dynamics. Finally, in Section IV the wave packets dynamics is investigated and a comparison with LQC showed. Concluding remarks follow.

Over the paper we adopt units such that $\hbar=c=1$.

\section{Snyder Space and Deformed Quantum Mechanics}

In this Section the modifications induced on the Heisenberg uncertainty relations by a non-commutative Snyder geometry are described. The Hilbert space representation and the deformed harmonic oscillator are also investigated. A relation with the $\kappa$-deformed Poincar\'e algebra is reviewed.

\subsection{Snyder-deformed Heisenberg algebras}

Let us start by considering a $n$-dimensional non-commutative (deformed) Euclidean space such that the commutator between the coordinates has the non-trivial structure ($\{i,j,...\}\in\{1,...,n\}$)
\be\label{snyalg}
[\tilde q_i,\tilde q_j]=\alpha M_{ij}\,,
\ee
where with $\tilde q_i$ we refer to the non-commutative coordinates and $\alpha\in\mathbb R$ is the deformation parameter with dimension of a squared length (for more details see \cite{BatMel08}). We then demand that the rotation generators $M_{ij}=-M_{ji}=i(q_ip_j-q_jp_i)$ satisfy the ordinary $SO(n)$ algebra and that the translation group is not deformed, i.e. $[p_i,p_j]=0$. We also assume that the rotational symmetry is preserved (the non-commutative coordinates transform as vectors under $SO(n)$ rotations), i.e. the commutators
\bea\label{commx}
[M_{ij},\tilde q_k]&=&\tilde q_i\delta_{jk}-\tilde q_j\delta_{ik}, \\\nonumber
[M_{ij},p_k]&=&p_i\delta_{jk}-p_j\delta_{ik}
\eea 
hold. This way we deal with the (Euclidean) Snyder space \cite{Sny}. The above relations however do not uniquely fix the commutators between $\tilde q_i$ and $p_j$. More precisely, there are infinitely many of such commutators which are all compatible (in the sense that the algebra closes in virtue of the Jacobi identities) with the above natural requirements. 

Such a feature can be understood considering a rescaling of the non-commutative coordinates $\tilde q_i$ in terms of ordinary phase space variables ($q_i,p_j$) \cite{Mel}. The most general $SO(n)$ covariant realization for $\tilde q_i$ is given by
\be\label{real}
\tilde q_i=q_i\varphi_1(\alpha p^2)+\alpha(q_jp_j)p_i\varphi_2(\alpha p^2),
\ee
where the convention $a_ib_i=\sum_i a_ib_i$ is adopted and $\varphi_1$ and $\varphi_2$ are two finite functions. The boundary condition we have to impose in order to recover the ordinary Heisenberg algebra as $\alpha=0$, reads $\varphi_1(0)=1$. The rescaling (\ref{real}) depends on the adopted algebraic structure, but the two functions $\varphi_1$ and $\varphi_2$ are not uniquely fixed. Given any function $\varphi_1$ satisfying the boundary condition $\varphi_1(0)=1$, the function $\varphi_2$ is thus determined by the relation \cite{BatMel08} $\varphi_2=(1+2\dot\varphi_1\varphi_1)/(\varphi_1-2\alpha p^2\dot\varphi_1)$ where $\dot\varphi_1=d\varphi_1/d(\alpha p^2)$. The commutator between $\tilde q_i$ and $p_j$ then arises from the realization (\ref{real}) and reads
\be\label{xpcom}
[\tilde q_i,p_j]=i\left(\delta_{ij}\varphi_1+\alpha p_ip_j\varphi_2\right).
\ee
From this relation we obtain the generalized uncertainty principle underlying the Snyder non-commutative space as
\be\label{unrel}
\Delta\tilde q_i\Delta p_j\geq\f12|\delta_{ij}\langle\varphi_1\rangle+\alpha\langle p_ip_j\varphi_2\rangle|
\ee
and the ordinary framework is recovered in the $\alpha\rightarrow0$ limit. Therefore, the deformation of the only commutator between the spatial coordinates as in (\ref{snyalg}) leads to infinitely many realizations of the algebra, and thus generalized uncertainty relations (\ref{unrel}), all of them consistent with the assumptions underlying the model. We also note that, unless $\varphi_2=0$, compatible observables no longer exist.

The most interesting feature to be stressed is that, for one-dimensional systems, this picture is almost uniquely fixed. In this case the symmetry group is trivial ($SO(1)=\text{Id}$) and the most general realization is given by $\tilde q=q\varphi(\alpha p^2)=q\sqrt{1-\alpha p^2}$. The commutation relation (\ref{xpcom}) reduces to 
\be\label{xp}
[\tilde q,p]=i\sqrt{1-\alpha p^2}
\ee
and the only freedom relies on the sign of the deformation parameter $\alpha$. It is worth noting that, when $\alpha>0$ a natural cut-off on the momentum arises, i.e. $|p|<\sqrt{1/\alpha}$, while as $\alpha<0$ the uncertainty relation (\ref{unrel}) predicts a minimal observable length $\Delta{\tilde q}_\text{min}=\sqrt{-\alpha}/2$. In fact, equation (\ref{unrel}) becomes
\be\label{uncrel}
\Delta\tilde q\Delta p\geq \f 1 2|\langle\sqrt{1-\alpha p^2}\rangle|,
\ee 
from which the minimal uncertainty in $\tilde q$, if $\alpha<0$, is obtained. Moreover, at the first order in $\alpha$, the string theory result \cite{String} $\Delta q\gtrsim(1/\Delta p+l_s^2\Delta p)$, in which the string length $l_s$ can be identified with $\sqrt{-\alpha/2}$, is recovered. On the other hand, if $\alpha>0$ a vanishing uncertainty in the non-commutative coordinate is allowed and appears as soon as $\Delta p$ reaches the critical value of $(\Delta p)^\star=\sqrt{(1-\alpha\langle p\rangle)/\alpha}$. We can then conclude that, a maximum momentum or a minimal length are predicted by the Snyder-deformed relation (\ref{xp}) if $\alpha>0$ or $\alpha<0$, respectively. 

\subsection{Hilbert space representation}

The Hilbert space representation of the deformed Heisenberg algebra (\ref{xp}) is constructed. Such a relation can be represented in the momentum space, where the $p$ and $\tilde q$ operators act on the wave function $\psi(p)=\langle p|\psi\rangle$ as
\be\label{rep}
p\,\psi(p)=p\,\psi(p), \qquad \tilde q\,\psi(p)=i\sqrt{1-\alpha p^2}\,\p_p\psi(p),
\ee
on a dense domain $D$ of smooth functions. Further, the self-adjoint requirement of the position and momentum operators implies a modified measure in the Hilbert space. In fact, $p$ and $\tilde q$ are self-adjoint operators in the domain $D$ with respect to the scalar product 
\be\label{scapro}
\langle\psi|\phi\rangle_\pm=\int_{I(\mathbb R)} \f{dp}{\sqrt{1-\alpha p^2}}\psi^\ast(p)\phi(p), 
\ee
where $I=\left(-1/\sqrt\alpha\,,1/\sqrt\alpha\right)$ and $\pm$ indicates the cases $\alpha>0$ and $\alpha<0$ respectively. Here the factor $(1-\alpha p^2)^{-1/2}$ in the measure is necessary in order to cancel the corresponding term of the operator representation of $\tilde q$. This way, the identity operators can be immediately obtained and the scalar product between momentum eigenstates appears modified as $\langle p|p'\rangle=\sqrt{1-\alpha p^2}\delta(p-p')$. The deformed Hilbert spaces $\mathcal F_\pm$, which are the Cauchy completions with respect to the inner product (\ref{scapro}), can be written as
\be
\mathcal F_\pm=L^2\left(I(\mathbb R),dp/\sqrt{1-\alpha p^2}\right). 
\ee
We note that these Hilbert spaces are unitarily inequivalent to each other and also with respect to the ordinary one $L^2(\mathbb R,dp)$ which appears as $\alpha\rightarrow0$. This is not surprising since the deformation of the canonical commutation relations can be viewed, from the realization (\ref{real}), as an algebra homomorphism which is a non-canonical transformation. In particular, it cannot be implemented at the quantum level as an unitary transformation. New features are then introduced at both classical and quantum level.

Let us now investigate how the position eigenvectors are modified in this framework. They satisfy, in the momentum space, the differential equation $i\sqrt{1-\alpha p^2}\,\p_p\psi_k(p)=k\psi_k(p)$ and explicitly read
\be\label{poseigen}
\psi^{(+)}_k(p)=c\exp\left(-i\f k{\sqrt\alpha}\sin^{-1}(\sqrt\alpha p)\right), \qquad \psi^{(-)}_k(p)=c\exp\left(-i\f k{\sqrt{-\alpha}}\sinh^{-1}(\sqrt{-\alpha} p)\right) 
\ee
$c$ being the normalization constant. These states generalize the plane waves and, as in the ordinary quantum mechanics, they are not normalizable. The $(+)$ and $(-)$ eigenstates can be obtained each other simply sending $\alpha\rightarrow-\alpha$, but an important difference between these states has to be stressed. As we have seen, if $\alpha<0$ a finite minimal uncertainty in position $\Delta{\tilde q}_\text{min}>0$ is predicted. This feature implies that there cannot be any physical state which is a position eigenstate since an eigenstate of an observable necessarily has vanishing uncertainty on it \cite{Kem}. To be more precise, in the ordinary quantum mechanics a sequence $\vert\psi_n\rangle\in D$ with position uncertainties decreasing to zero, exists. On the other hand, in presence of a minimal uncertainty, it is not longer possible to find some $\vert\psi_n\rangle\in D$ such that 
\be
\lim_{n\rightarrow\infty}\left(\Delta \tilde q\right)_{|\psi_n\rangle}=\lim_{n\rightarrow\infty}\langle\psi|(\tilde q-\langle\psi|\tilde q|\psi\rangle)^2|\psi\rangle=0.
\ee
Thus, although it is possible to construct position eigenvectors, they are only formal eigenvectors and not physical states. In this case we have lost direct information on the position itself (it can be recovered by an analysis like the one in \cite{Kem}). On the other hand, when $\alpha>0$ the position eigenvectors $\psi^{(+)}_k(p)$ in (\ref{poseigen}) are ``proper physical states'' in the sense of the standard quantum theory (a zero uncertainty in position is allowed), but two remarks are in order. (i) They are generally no longer orthogonal\footnote{The position operator is no longer essentially self-adjoint but has a one-parameter family of self-adjoint extensions. Considering $k=\sqrt\alpha(2n+\lambda)$ $(n\in\mathbb Z,\,\lambda\in(-1,1))$, a one-parameter family of eigenvectors $\psi^{(+)}_k$ which explicitly diagonalizes the position operator is constructed and a lattice spacing $2\sqrt\alpha$ is then introduced.}. In fact, the scalar product of these eigenvectors appear to be $\langle\psi^{(+)}_{k'}|\psi^{(+)}_k\rangle_-=2c^2\sqrt{\alpha}(\pi(k-k'))^{-1} \sin\left(\pi(k-k')/2\sqrt\alpha\right)$, instead of the Dirac $\delta$-distribution. (ii) They have a finite energy, namely the mean value of $p^2$ between such states is finite, i.e. $\langle\psi^{(+)}_k|p^2|\psi^{(+)}_k\rangle_-=c^2\pi/2\alpha^{3/2}$. Of course, when the limit $\alpha\rightarrow0$ is taken into account, the usual results of Heisenberg quantum mechanics are recovered. 

\subsection{The harmonic oscillator}

We now investigate a direct physical prediction of the framework discussed above. The harmonic oscillator is surely one of the most relevant mechanical systems for testing any quantization scheme and therefore we apply such a formalism to this model, focusing on the modifications on the energy spectrum induced by the deformation parameter $\alpha$. Considering the classical Hamiltonian with a Snyder-deformed quadratic potential
\be
\mathcal H=\f{p^2}{2m}+\f12m\omega^2\tilde q^2
\ee
and the representation for $p$ and $\tilde q$ as reported in (\ref{rep}), we immediately get the deformed stationary Schr\"odinger equation for the model
\be\label{modho}
\psi''(p)-\f{\alpha p}{1-\alpha p^2}\psi'(p)+\f1{1-\alpha p^2}\left(\epsilon-d^4p^2\right)\psi(p)=0,
\ee
where the prime denotes differentiation with respect to $p$ and we have defined $\epsilon=2E/m\omega^2$ and the characteristic length scale $d=1/\sqrt{m\omega}$. We note that if $\alpha>0$, i.e. $p\in I$, no singularities appear in (\ref{modho}). This equation is well-known in mathematics as the so-called Mathieu equation in the algebraic form \cite{AS}. Its solution can be explicitly written in terms of the Mathieu cosine $\mathcal C$ and sine $\mathcal S$ as 
\be
\psi(p)=A\mathcal C\left(\nu,q,\cos^{-1}(\sqrt{-\alpha} p)\right)+B\mathcal S\left(\nu,q,\cos^{-1}(\sqrt{-\alpha} p)\right),
\ee
where $A, B$ are integration constants and
\be\label{defnuq}
\nu=\f{-d^4-2\alpha\epsilon}{2\alpha^2}, \qquad q=\f{d^4}{4\alpha^2}.
\ee   
Modifications on the ordinary energy eigenvalues $E=E_n$, induced by the deformed algebra, can be easily obtained considering an asymptotic formula for the $\nu$ coefficients. In fact, we are interested only at first-order corrections to the spectrum and these appear for $\alpha\rightarrow0$, or more precisely when the scale $d^2$ of the harmonic oscillator is much bigger then the deformation scale $\alpha$, i.e. when $\alpha/d^2\ll1$. In other words, we are interested when $q\gg1$ and in this case $\nu$ can be expanded as \cite{AS}
\be
\nu=\nu_n=-2q+2\sqrt q(2n+1)-\f{2n^2+2n+1}4+\mathcal O\left(\f1{\sqrt q}\right).
\ee 
This way, considering the definitions (\ref{defnuq}), the deformed energy spectrum reads
\be\label{defspec}
E_n=\f\omega2(2n+1)-\f\omega8(2n^2+2n+1)\left(\f\alpha{d^2}\right)+\mathcal O\left(\f{\alpha^2}{d^4}\right)
\ee 
and, as expected, for $\alpha/d^2\rightarrow0$ the ordinary eigenvalues are recovered. Let us discuss such a result. The spectrum in the $\alpha>0$ case is nothing but the one obtained in the polymer quantum mechanics \cite{pol}. This framework relies on a non-standard representation of the canonical commutation relations (inspired by loop quantum gravity \cite{LQG}) which is (unitarily) inequivalent to the Heisenberg one. A notable fact is that such type of quantization when applied to the minisuperspace models, leads to LQC \cite{LQC}. Differently, the spectrum (\ref{defspec}) for $\alpha<0$ appears to be the same as the one achieved in minimal length quantum mechanics \cite{Kem}, i.e. considering the fundamental commutator as $[\tilde q,p]=i(1+\beta p^2)$, which can be considered as the first order approximation in $\alpha=-2\beta$ of (\ref{xp}).

\subsection{Relation with the $\kappa$-Poincar\'e algebra}

The non-commutative Snyder space has been analyzed in the literature from different points of view \cite{Kov,LO,tt}, but only two particular realizations of its algebra are known: the Snyder \cite{Sny} and the Maggiore \cite{Mag} ones. The original realization of Snyder is recovered as a special case of (\ref{real}) if $\varphi_1=1$. On the other hand, the Maggiore algebra $[\tilde q_i,p_j]=i\delta_{ij}\sqrt{1-\alpha p^2}$, can be regarded as the particular case of (\ref{real}) when the condition $\varphi_2=0$ is taken into account. In this case the Snyder framework can be related to the $\kappa$-Poincar\'e scheme in the following sense \cite{Mag}. The $\kappa$-Poincar\'e algebra (for reviews see \cite{kP}) provides an explicit realization of the Snyder-deformed algebra, once some identifications are taken into account. If $\tilde q$ is identified with a suitably $\kappa$-deformed Newton-Wigner position operator and $p$ and $M_{ij}$ as the generators of spatial translations and rotations of the $\kappa$-Poincar\'e algebra respectively, the modified Heisenberg relations are recovered. (For other comparisons between deformed Heisenberg algebras and $\kappa$-Poincar\'e see \cite{gupk}.) Physical interest in the $\kappa$-Poincar\'e algebra arises since it is the mathematical structure of DSR. Moreover, this framework describes the symmetries of fields living on $\kappa$-Minkowski non-commutative space-times and is widely expected that the study of (quantum) fields, invariant under such symmetries, may give physical insights on a flat space limit of quantum gravity \cite{Arz}.

\section{Deformed Dynamics of the FRW models}

We investigate the Snyder-deformed dynamics of the isotropic cosmological models. The system is studied at classical level searching for the modifications induced by the deformed Heisenberg algebra. We start reviewing the ordinary FRW dynamics and then turn to the deformed one. 

\subsection{Ordinary canonical dynamics}

The FRW cosmological models are (spatially) isotropic models described by the line element 
\be
ds^2=-N^2dt^2+a^2\left(\f{dr^2}{1-kr^2}+r^2d\Omega^2\right),
\ee
where $N=N(t)$ is the lapse function and $a=a(t)$ the scale factor. The lapse function does not play a dynamical role while the scale factor is the only degree of freedom of the system describing the expansion of the Universe. The parameter $k$ can be zero or $\pm1$ depending on the symmetry group. The dynamics of such models is summarized in the scalar constraint
\be\label{scacon}
\mathcal H=-\f{2\pi G}3\f{p_a^2}a-\f3{8\pi G}ak+a^3\rho=0,
\ee
where $G=l_P^2$ in the gravitational constant and $\rho=\rho(a)$ denotes a generic energy density we have introduced into the system. Therefore, isotropy reduces the phase space of general relativity to be 2-dimensional in which the only non-vanishing Poisson bracket is $\{a,p_a\}=1$. The Friedmann equation can be extracted by using the Hamilton equations with respect to the extended Hamiltonian
\be\label{extham}
\mathcal H_E=\f{2\pi G}3N\f{p_a^2}a+\f3{8\pi G}Nak-Na^3\rho+\lambda\pi,
\ee
where $\lambda$ is a Lagrange multiplier and the term $\lambda\pi$ is introduced since $\pi$, the momenta conjugate to $N$, vanishes. We note that $\dot N=\{N,\mathcal H_E\}=\lambda$ and that the scalar constraint (\ref{scacon}) is obtained requiring the constraint $\pi=0$ will be satisfied at all times, i.e. demanding that the secondary constraint $\dot\pi=\{\pi,\mathcal H_E\}=\mathcal H=0$ holds. The remaining equations of motion with respect to $\mathcal H_E$ read
\be\label{eqap}
\dot a=\{a,\mathcal H_E\}=\f{4\pi G}3N\f{p_a}a, \qquad \dot p_a=\{p_a,\mathcal H_E\}=N\left(\f{2\pi G}3\f{p_a^2}{a^2}-\f3{8\pi G}k+3a^2\rho+a^3\f{d\rho}{da}\right).
\ee
Making use of the above equations and the scalar constraint (\ref{scacon}), we immediately obtain the equation of motion for the Hubble rate $(\dot a/a)$ as  
\be\label{canfri}
\left(\f{\dot a}a\right)^2=\f{8\pi G}3\rho-\f k{a^2},
\ee
which is the desired Friedmann equation in a synchronous reference frame\footnote{In the synchronous reference frame, defined in the $3+1$ framework by $N=1$ and $N^i=0$, the time coordinate identifies with the proper time at each point of space.}. It is well-known that this equation leads to the big-bang singularity where the (general-relativistic) description of the Universe is no longer appropriate and quantum modifications are required.  

\subsection{Deformed canonical dynamics}   

We now perform the analysis of the deformed dynamics of the FRW models and therefore we consider the one-dimensional case of the scheme analyzed above. More precisely, we check the modifications arising from the algebra (\ref{xp}) on the classical trajectory of the Universe described in the previous Section. A quantum cosmological bouncing solution is obtained and it resembles the one achieved in recent issues of LQC (if $\alpha>0$). As $\alpha<0$ the Randall-Sundrum braneworld scenario is recovered. 

The Snyder-deformed classical dynamics is summarized in the modified symplectic geometry arising from the classical limit of (\ref{xp}), as soon as the parameter $\alpha$ is regarded as an independent constant with respect to $\hbar$. It is then possible to replace the quantum-mechanical commutator (\ref{xp}) via the Poisson bracket
\be\label{pm}
-i[\tilde q,p]\Longrightarrow\{\tilde q,p\}=\sqrt{1-\alpha p^2}.
\ee
We stress once again that this relation corresponds exactly to the unique (up to a sign) possible realization of the Snyder space. In order to obtain the deformed Poisson bracket, some natural requirements have to be considered. As a matter of fact, it must possess the same properties as the quantum mechanical commutator, i.e. it has to be anti-symmetric, bilinear and satisfy the Leibniz rules as well as the Jacobi identity. This way, the Poisson bracket (for any two-dimensional phase space function) appears to be
\be
\{F,G\}=\left(\f{\p F}{\p\tilde q}\f{\p G}{\p p}-\f{\p F}{\p p}\f{\p G}{\p\tilde q}\right)\sqrt{1-\alpha p^2}.
\ee
In particular, the time evolution of the coordinate and momentum with respect to a given deformed Hamiltonian $\mathcal H(\tilde q,p)$, are now specified by
\be
\dot{\tilde q}=\{\tilde q,\mathcal H\}=\f{\p\mathcal H}{\p p}\sqrt{1-\alpha p^2}, \qquad \dot p=\{p,\mathcal H\}=-\f{\p\mathcal H}{\p\tilde q}\sqrt{1-\alpha p^2}.
\ee
Let us apply this framework to the FRW model in the presence of a generic matter energy density, namely to the Hamiltonian (\ref{extham}). Therefore we assume the minisuperspace as Snyder-deformed and then the commutator between the isotropic scale factor $a$ and its conjugate momentum $p_a$ is uniquely fixed by the relation 
\be\label{ap}
\{a,p_a\}=\sqrt{1-\alpha p_a^2}\,,
\ee
while the equations of motion $\dot N=\{N,\mathcal H_E\}=\lambda$ and $\dot\pi=\{\pi,\mathcal H_E\}=\mathcal H=0$ remain unchanged. In fact, the Poisson bracket $\{N,\pi\}=1$ is not affected by the deformations induced by the $\alpha$ parameter on the system. On the other hand, the equations of motion (\ref{eqap}) become modified in such an approach via the relation (\ref{ap}) and read
\be\label{eqapgup}
\dot a=\{a,\mathcal H_E\}=\f{4\pi G}3N\f{p_a}a\sqrt{1-\alpha p_a^2}, \qquad \dot p_a=\{p_a,\mathcal H_E\}=N\left(\f{2\pi G}3\f{p_a^2}{a^2}-\f3{8\pi G}k+3a^2\rho+a^3\f{d\rho}{da}\right)\sqrt{1-\alpha p_a^2}.
\ee 
As in the canonical case, the equation of motion for the Hubble rate can be obtained solving the constraint (\ref{scacon}) with respect to $p_a$ and then considering the first equation of (\ref{eqapgup}). Explicitly it becomes (taking $N=1$)
\be\label{deffri}
\left(\f{\dot a}a\right)^2=\left(\f{8\pi G}3\rho-\f k{a^2}\right)\left[1-\f{3\alpha}{2\pi G}a^2\left(a^2\rho-\f3{8\pi G}k\right)\right].
\ee
We refer to this equation as the {\it deformed Friedmann equation} as it entails the modification arising from the Snyder-deformed Heisenberg algebra (\ref{xp}). It is interesting to consider the flat FRW Universe, i.e. the $k=0$ model. In this case the deformed equation (\ref{deffri}) appears to be
\be\label{modfri}
\left(\f{\dot a}a\right)^2_{k=0}=\f{8\pi G}3\rho\left(1-\text{sgn}\,\alpha\f\rho{\rho_c}\right),
\ee
where $\rho_c=(2\pi G/3|\alpha|)\rho_P$ is the critical energy density, $\rho_P$ being the Planck one. In the last step we have assumed the existence of a fundamental minimal length. In fact, as widely accepted, one of the most peculiar consequences of all promising quantum gravity theories is the existence of a fundamental cut-off length, which should be related to the Planck one (for a review see \cite{gar}). Therefore, although this minimal length appears differently in distinct contexts, it is reasonable that the scale factor (the energy density) has a minimum (maximum) at the Planck scale.

The modifications arising from the deformed Heisenberg algebra on the Friedmann equation (\ref{modfri}) are manifested in the form of a $\rho^2$-term. Such factor is relevant in high energy regime and, if $\alpha>0$ and $\rho$ reaches the critical value $\rho_c$, the Hubble rate vanishes and the Universe experiences a bounce (or more generally a turn-around) in the scale factor. For energy density much smaller then the critical one the standard Friedmann dynamics, equation (\ref{canfri}) for $k=0$, is recovered. In the same way, when the deformation parameter $\alpha$ vanishes, the correction term disappears and the ordinary behavior of the Hubble parameter is obtained. 

The interesting feature to be stressed is the equivalence, at phenomenological level, between the deformed Friedmann equation (\ref{modfri}) in the $\alpha>0$ case and the one obtained considering the effective dynamics of LQC \cite{elqc}. On the other hand, the string inspired Randall-Sundrum braneworld scenario leads to a modified Friedmann equation as in (\ref{modfri}) with $\alpha<0$ \cite{Roy}. The opposite sign of the $\rho^2$-term in such an equation, is the well-known key difference between the loop cosmology and the Randall-Sundrum framework. In fact, the former approach leads to a non-singular bouncing cosmology while in the latter, because of the positive sign, $\dot a$ cannot vanish and a cosmological bounce cannot take place. Of course, to obtain a bounce the correction should be negative, i.e. make a repulsive contribution. This can occur also in the Randall-Sundrum scheme as soon as the extra dimension of the bulk space-time is considered to be time-like \cite{sha1}. However, the minus sign in this approach remains an open question and no definitive answers are given \cite{sha2}.

\section{Physical Considerations on the Model}

A peculiar model is investigated in the framework of the Snyder-deformed minisuperspace with particular attention to the evolution of the relative fluctuations of the scale factor. Comparison with other similar approaches is then showed.

\subsection{Flat isotropic model filled with a scalar field}

An isotropic flat Universe filled with a massless scalar field $\phi$ is analyzed in the context of the previous discussion. Such a model deserves interest since it is the one most studied in the framework of LQC. The energy density of this scalar field takes the form $\rho=p_\phi^2/2a^6$, where $p_\phi$ denotes the momentum canonically conjugate to $\phi$, and then the scalar constrain (\ref{scacon}) becomes
\be\label{conphi}
\mathcal H=-\f{2\pi G}3\f{p_a^2}a+\f{p_\phi^2}{2a^3}=0.
\ee
The phase space is $4$-dimensional, with coordinates $(a,p_a,\phi,p_\phi)$ and, since $p_{\phi}$ is a constant of motion, each classical trajectory is specified in the $(a,\phi)$-plane. The scalar field $\phi$ can be then regarded as an internal clock for the dynamics and this condition can be imposed requiring the time gauge $\dot\phi=1$, i.e. $N=a^3/p_\phi$. In this case the deformed Friedmann equation (\ref{deffri}) rewrites as
\be
\left(\f{\dot a}a\right)^2=\f{4\pi G}3\left(1-\f{3\alpha}{4\pi G}\f{p_\phi^2}{a^2}\right),
\ee
whose solution is given by $a(\phi)\sim e^{-\sqrt{4\pi G/3}\phi}(\alpha\tilde p_\phi^2+e^{2\sqrt{4\pi G/3}\phi})$, where $\tilde p_\phi^2=9p_\phi^2/16\pi^4 G^2$. This equation clearly predicts a big-bounce if $\alpha>0$ (from now on we consider only this case). The ordinary solutions $a(\phi)\sim e^{\pm\sqrt{4\pi G/3}\phi}$ are recovered at late times, i.e. at $|\phi|\rightarrow\infty$. Fixing the lapse function as before, the (effective) Hamiltonian in the internal time $\phi$ description is given by $H=\sqrt{4\pi G/3}\,p_aa$. The time evolution of any observable $\mathcal O$ can then be realized with respect to such a Hamiltonian, i.e. the equation of motion for the expectation value $d\langle\mathcal O\rangle/d\phi=-i\langle[\mathcal O,H]\rangle$ holds. The equations of motion 
\be
\f d{d\phi}\langle a\rangle=\sqrt{\f{4\pi G}3}\left\langle a\sqrt{1-\f{3\alpha}{4\pi G}\f{p_\phi^2}{a^2}}\right\rangle, \qquad 
\f d{d\phi}\langle p_a\rangle=-\sqrt{\f{4\pi G}3}\left\langle p_a\sqrt{1-\f{3\alpha}{4\pi G}\f{p_\phi^2}{a^2}}\right\rangle
\ee
are thus immediately obtained and these trajectories are in exact agreement with the classical ones. We are now interested to investigate the semi-classical proprieties of such a Snyder-deformed quantum Universe. To be precise, with semi-classical requirement for an observable $\mathcal O$ we refer to the requirement that its expectation value be close to the classical one and that the relative fluctuations $(\Delta\mathcal O)^2/\langle\mathcal O\rangle^2\ll1$. We now analyze the evolution of the scale factor relative fluctuations which are governed by the equation 
\be\label{relflu}
\f d{d\phi}\left(\f{(\Delta a)^2}{\langle a\rangle^2}\right)=\sqrt{\f{16\pi G}3}\f1{\langle a\rangle^2}\left(\left\langle a^2\sqrt{1-\f{3\alpha}{4\pi G}\f{p_\phi^2}{a^2}}\right\rangle-\f{\langle a^2\rangle}{\langle a\rangle}\left\langle a\sqrt{1-\f{3\alpha}{4\pi G}\f{p_\phi^2}{a^2}}\right\rangle\right).
\ee
As we can see, in the ordinary framework ($\alpha=0$), such a quantity is conserved during the whole evolution (the only fluctuations $(\Delta a)^2$ are not constants and neither bounded) and thus the semi-classicity of an initial state is preserved. Such a propriety is also valid in the deformed scheme at late times $|\phi|\rightarrow\infty$, i.e. for large scale factor $a\gg\sqrt{\alpha}p_\phi/l_P$. We note that at the bouncing time, i.e. when the scale factor reaches its minimum value $a_\text{min}=\sqrt{3\alpha/4\pi G}\,p_\phi$, the derivative of this uncertainty vanishes.    

Let us consider the deformed Wheeler-DeWitt (WDW) equation for this model. Regarding the massless scalar field $\phi$ as a relational time for the evolution and using the representation (\ref{rep}), the constraint (\ref{conphi}) takes the form
\be\label{wdweq}
\p_\phi^2\Psi(p_a,\phi)=-\Theta\Psi(p_a,\phi), \qquad \Theta=-\f{4\pi G}3p_a^2\left((1-\alpha p_a^2)\p_{p_a}^2-\alpha p_a\p_{p_a}\right).
\ee
As usual the WDW equation has the same form as the Klein-Gordon equation where $\Theta$ plays the role of the ordinary Laplacian. In order to have an explicit Hilbert space, we perform the natural frequencies decomposition of the solution of (\ref{wdweq}) and focus on the positive frequency sector. The wave function $\Psi_\omega(p_a,\phi)=e^{i\omega\phi}\psi_\omega(p_a)$ is thus of positive frequency with respect to $\phi$ and satisfies the positive frequency (square root) of the quantum constraint (\ref{wdweq}), i.e. we deal with a Sch\"odinger-like equation $-i\p_\phi\Psi=\sqrt\Theta\Psi$ with a non-local Hamiltonian $\sqrt\Theta$ (here $\omega^2$ denotes the spectrum of $\Theta$ and covers the interval $(0,\pi G/3)$). The wave function $\psi_\omega$ is explicitly expressed in terms of the Gauss hypergeometric functions $F$ and reads
\be
\psi_\omega(p_a)=Ap_a^{(1-\gamma)/2}F\left(\f14(1-\gamma),\f14(1-\gamma),1-\f\gamma2;\alpha p_a^2\right)+Bp_a^{(1+\gamma)/2}F\left(\f14(1+\gamma),\f14(1+\gamma),1+\f\gamma2;\alpha p_a^2\right),
\ee
where $\gamma=\sqrt{1-3\omega^2/\pi G}$. It is worth stressing that this function is well defined (is not divergent) since we remember the existence of a cut-off on the momentum, i.e. $p_a\in I$. This way, taking a weighting function $A(\omega)$, we can construct a wave packet which has the general form
\be\label{wavpac}
\Psi(p_a,\phi)=\int d\omega A(\omega)\psi_\omega(p_a)e^{i\omega\phi}.
\ee
As last step of our analysis we note that, although the relative fluctuations $(\Delta a)^2/\langle a\rangle^2$ are in general not constant during the evolution (equation (\ref{relflu})), the difference in the asymptotic values 
\be
D=\lim_{\phi\rightarrow\infty}\left|\left(\f{(\Delta a)^2}{\langle a\rangle^2}\right)_{-\phi}-\left(\f{(\Delta a)^2}{\langle a\rangle^2}\right)_\phi\right|
\ee
vanishes. This consideration can be realized since the fluctuations $(\Delta a)^2(\phi)$ and the mean value $\langle a\rangle(\phi)$ are symmetric in time. In fact, given any real $A(\omega)$ (for example a Gaussian weighting function $e^{-(\omega-\omega_0)^2/2\sigma^2}$), the mean value of any self-adjoint operator $\mathcal O$ with respect to the states (\ref{wavpac}) is invariant under time inversion $\phi\rightarrow-\phi$. Therefore, starting with a Gaussian semi-classical state such that $(\Delta a)^2/\langle a\rangle^2\ll1$ at late times, this propriety is satisfied on the other side of the bounce when the Universe approaches large scale ($a\gg\sqrt{\alpha}p_\phi/l_P$).  

\subsection{Comparison with other approaches}

Our model can be regarded as an attempt to mimic the original LQC system by a simpler one and in this respect it has to be compared with analogous existing approaches. There are essentially two (related) ways to capture the essential features of the original quantum system by an exactly solvable one and both these frameworks regard the cosmological model described by (\ref{conphi}). The first approach \cite{Boj1} is based on replacing the connection by its sine (the connection itself cannot be directly implemented as an operator in the loop Hilbert space) and then rewriting the Hamiltonian in term of non-canonical variables. This model serves as a perturbative basis for realistic bounce scenarios and allows a precise analysis of the evolution of dynamical coherent states. The second approach \cite{Cor} relies on a reduction of the so-called improved dynamics and allows to clearly define in which sense the WDW theory approximates LQC. Both models agree, although this is matter of current debate \cite{Boj2}, with the claim that semi-classicality is preserved across the cosmological quantum bounce \cite{BC}. This feature is in agreement with our approach. However two remarks are in order. (i) The (simplified) LQC theory is based on a Weyl representation of the canonical commutation relations which turns out to be inequivalent to the Sch\"odinger representation. On the other hand, as explained before, the Snyder-deformed algebra cannot be obtained by a canonical transformation of the ordinary Poisson brackets of the system. (ii)\footnote{I thank Martin Bojowald for stressing me this point.} The $\rho^2$-term in the modified Friedmann equation (\ref{modfri}) is not the only correction from LQC unless the only matter source is a massless scalar field. If it has mass or is self-interacting, there are infinitely many other correction terms which also involve pressure \cite{Boj3}. In our model the form of (\ref{modfri}) is independent of the precise matter content. 

\section{Concluding remarks}

In this paper we have shown that a bouncing cosmology is predicted by a Snyder-deformed Friedmann dynamics. In particular, we have implemented a Snyder non-commutative geometry, which can be related to DSR as well as to the $\kappa$-Poincar\'e algebra, in the FRW minisuperspace arena. Since we deal with a one-dimensional system, the deformation is almost uniquely fixed and a cosmological bounce is then obtained. Our deformed Friedmann equation has the same form of the LQC one. Notably, also the effective cosmological dynamics of the Randall-Sundrum braneworld scenario is allowed because of the freedom in the sign of the deformation term. Such a result is also corroborated by the analysis of the Snyder-deformed harmonic oscillator. The LQC-like framework is the one in which a cut-off on the momentum is predicted. On the other hand, in the braneworld-like one, a minimal observable length arises and the string theory uncertainty relation is recovered. Summarizing, a non-commutative (deformed) picture which leads, at phenomenological level, to the predictions of more general theories can be formulated. The validity of such an approach in more general, and physically interesting, frameworks will be subject of future investigations. 

\bigskip

{\bf Acknowledgments.} I thank Giovanni Montani for having encouraged this work. Francesco Cianfrani and Orchidea Maria Lecian are thanked for several discussions and for a critical reading of the manuscript.

\end{document}